\documentclass[twocolumn,amsmath,amssymb,pra]{revtex4-2}
\usepackage{txfonts}
\usepackage{graphicx}% Include figure files\underline{}
\usepackage{dcolumn}% Align table columns on decimal point
\usepackage{bm}% bold math
\usepackage{color}
\usepackage{graphicx}
\usepackage{braket}

\usepackage[colorlinks=true,	linkcolor=blue,urlcolor=blue,anchorcolor=blue,citecolor=blue,bookmarksnumbered]{hyperref}

\usepackage{xr}
\usepackage{nicematrix} % for boxed blocks
\usepackage{tikz}       % needed for drawing

\usepackage{mathrsfs}
\usepackage{bbm}
\usepackage{mathtools}

\begin{document}
\title{\textbf{Engineer coherent oscillatory modes in Markovian open quantum systems} 
}% 

\author{Chun Hei Leung$^1$}
\author{Pak-Tik Fong$^2$}
\author{Tianyi Yan$^1$}
\author{Weibin Li$^1$}
\affiliation{
$^1$School of Physics and Astronomy and Centre for the Mathematics and Theoretical Physics of Quantum Non-equilibrium Systems, University of Nottingham, Nottingham NG7 2RD, United Kingdom}
\affiliation{
$^2$Department of Physics, Simon Fraser University, Burnaby, British Columbia V5A 1S6, Canada}

\date{\today}% It is always \today, today,

\begin{abstract}

   We develop a novel framework to engineer persistent oscillatory modes in Markovian open quantum systems governed by a time-independent Lindblad master equation. We show that oscillatory modes can be created when the Hamiltonian and jump operator can be expressed in the same block-diagonal form. A key feature of the framework is that the dissipator of the Lindblad master equation are generally non-zero. We identify the weak and strong conditions, where the onset of the oscillatory modes is dependent and independent of the parameters of the system, respectively. Our method extends beyond the typical decoherence-free subspace approach, in which the dissipator is zero. We demonstrate the applicability of this framework using various models, showing how carefully tailored system-environment interactions can produce sustained coherent oscillations. 
  
% \begin{description}
% Secondary publications and information retrieval purposes.
% You may use the \texttt{description} environment to structure your abstract;
% use the optional argument of the \verb+\item+ command to give the category of each item. 
% \end{description}
\end{abstract}

\maketitle

%\tableofcontents

\section{\label{sec: Introduction} Introduction}

    Understanding the dynamics of open quantum systems is central to modern quantum science, such as solid-state physics \cite{simoni2025firstprinciplesopenquantumdynamics}, quantum thermodynamics \cite{Pathania2025}, quantum biology \cite{doi:10.1073/pnas.1005484107, Plenio_2008}, and quantum chemistry \cite{Kapral_2015, Oz2023, simoni2025firstprinciplesopenquantumdynamics}. Insights from the understanding allow to applications in quantum control~\cite{Koch_2016, PhysRevLett.121.035301, PhysRevResearch.2.023026, WEIDNER2025111987}, quantum sensing~\cite{Montenegro2023, jiao2024observationtimecrystalcomb, PhysRevLett.132.050801, PhysRevA.108.022413}, and quantum computing~\cite{PhysRevA.57.3276, PhysRevA.111.033718}. 
    System-environment coupling typically induces dissipation. As a result, the system equilibrates to stationary states, while oscillations are suppressed in the long time limit~\cite{Spohn1977, Menczel_2019}. Under certain conditions, however, persistent oscillations can survive indefinitely despite continuous environmental coupling \cite{bucaNonstationaryCoherentQuantum2019, albertSymmetriesConservedQuantities2014, Booker_2020, PhysRevLett.132.150401, Buča_2012, Zhang_2020, PhysRevLett.134.050407, Menczel_2019, PRXQuantum.5.030325, PhysRevA.77.052301}. Persistent oscillations have been investigated in various models, ranging from collective dissipative spin ensembles \cite{PhysRevA.107.032219, PhysRevB.102.174309}, synchronization of quantum oscillators \cite{PhysRevA.100.022119}, dissipative time crystals \cite{ PhysRevB.106.224308, li2025symmetryinducedfragmentationdissipativetime} to quantum scar in thermalized many-body systems \cite{Serbyn2021}.  Markovian open quantum systems are typically modeled using the Lindblad master equation \cite{Lindblad1976OnTG, 10.1063/1.522979, PhysRevB.102.115109}. Existing theoretical frameworks based on this description have successfully linked the emergence of persistent oscillations to the presence of decoherence-free states \cite{bucaNonstationaryCoherentQuantum2019, albertSymmetriesConservedQuantities2014, Booker_2020, PhysRevLett.132.150401, Buča_2012, Zhang_2020}. In these approaches, the symmetries of the system are identified, which protect the coherent evolution of non-stationary states. One framework requires the existence of decoherence-free subspaces or dark states of the jump operators~ \cite{PhysRevLett.81.2594, PhysRevLett.132.150401, Buča_2012, Zhang_2020}, such that the dissipation of states within such a subspace is zero. However, this requires that all states within the subspace be degenerate eigenstates of every jump operator, which is a highly restrictive condition for many practical implementations. Another approach depends on identifying an implicit eigenoperator or symmetry operator of the Liouvillian \cite{bucaNonstationaryCoherentQuantum2019, albertSymmetriesConservedQuantities2014, Booker_2020}. It predicts a more general condition than the decoherence-free subspace framework because it does not require the persistent oscillation state to be non-dissipative. On the other hand, such operators are not always straightforward to construct, particularly in complex systems. In addition, its connection to decoherence-free subspaces is less transparent.
    Despite these advances, the search for persistent oscillatory modes in open quantum systems has attracted growing attention. 

    In this work, we develop a general framework for constructing persistent oscillatory dynamics in Markovian open quantum systems with a time-independent Hamiltonian. The key idea here is that both the jump operator and Hamiltonian can be written in the same block-diagonal form with the same basis. Then, the multiple eigenvalues of the Liouvillian lie on the imaginary axis. Persistent oscillatory modes emerge in the dynamics when initial states overlap with the modes that have non-zero imaginary parts. Our approach not only recovers the connection to the decoherence-free subspace but also shows that persistent oscillations can occur without the existence of decoherence-free states, and can be present when the dissipators do not vanish. Beyond its fundamental significance, the insights gained from this approach open routes to exploit non-stationary quantum technologies, such as autonomous quantum clocks \cite{PhysRevX.7.031022, Milburn02042020, PhysRevX.11.021029}, by enabling a more flexible and constructive method for generating persistent oscillations and broadening the class of systems capable of supporting such dynamics in the future.
    
    The remainder of this paper is organized as follows. Sec.~\ref{sec: Background} reviews the Lindblad master equation and shows that the dynamical equation supports persistent oscillations. Sec.~\ref{sec: DFS} shows the framework connecting persistent oscillations with the decoherence-free subspace. Our generalization to the decoherence-free subspace framework is presented in Sec.~\ref{sec: general condition}, and its physical interpretation is discussed in Sec.~\ref{sec: Physical Interpretation}. In Sec.~\ref{sec: Examples}, we illustrate both frameworks using four examples. The oscillatory behavior observed in the first two examples (Sec.~\ref{sec: Example 0} and \ref{sec: Example 1}) is captured by the decoherence-free subspace framework, whereas the oscillations in the latter two examples (Sec.~\ref{sec: Example 2} and \ref{sec: Example 3}) can only be explained by our generalized framework but not by the decoherence-free subspace framework. Finally, Sec.~\ref{sec: Summary} concludes the paper.

%\clearpage
\section{\label{sec: Background} Lindblad master equation and Liouvillian eigenspectrum}

    %We first briefly review the Lindblad master equation and its eigenstates and eigenspectra. 
    We consider the Markovian Lindblad master equation \cite{Lindblad1976OnTG, 10.1063/1.522979, PhysRevB.102.115109}, which describes the evolution of a density matrix $\hat{\rho}$, 
    \begin{eqnarray}\label{eq: master equation}
        \frac{d}{dt} \hat{\rho} = \mathscr{L} [\hat{\rho}] 
            \equiv -\mathbbm{i} [ \hat{H} , \hat{\rho} ] + \sum_i \gamma_i \, \mathcal{D}_i [\hat{\rho}],
    \end{eqnarray}
    where $\mathscr{L}$ is the Liouvillian superoperator, $\hat{H}$ is the system Hamiltonian, and $\mathcal{D}_i [\hat{\rho}] = \hat{L}_i \hat{\rho} \hat{L}^\dag_i - \frac{1}{2} \hat{L}^\dag_i \hat{L}_i \hat{\rho} - \frac{1}{2} \hat{\rho} \hat{L}^\dag_i \hat{L}_i$ is the $i$-th dissipator with the jump operator $\hat{L}_i$ with damping rate $\gamma_i$. 
    When $\mathscr{L}$ is diagonalizable, the solution of Eq.~\eqref{eq: master equation} can be expressed \cite{PRXQuantum.5.030325, PhysRevA.98.042118} as,
    \begin{equation}
        \hat{\rho} (t) = \sum_{k} c_k \, e^{\Lambda_k t} \,\hat{r}_k ,
    \label{eq: eigenmodes expansion}
    \end{equation}
    where $c_k = \operatorname{Tr} [\hat{l}^\dagger_k \hat{\rho} (0)] $ is the initial projection amplitude onto the $k$-th (right) eigenmode. $\Lambda_k$, $\hat{r}_k$, and $\hat{l}_k$ are the Liouvillian eigenvalues, right eigenmodes, and left eigenmodes of the Liouvillian superoperator $\mathscr{L}$~\cite{PhysRevResearch.5.043036,longhiQuantumMpembaEffect2025,zhangObservationQuantumStrong2025}, respectively. 
    It can be shown that the real part of the Liouvillian eigenvalues is non-positive \cite{zhang2025directalgebraicproofnonpositivity, Baumgartner_2008}.
    Importantly, there always exists an eigenmode $r_\text{SS}$ with a zero Liouvillian eigenvalue, $\mathscr{L} [ \hat{r}_\text{SS} ] = 0$, indicating that a steady mode always exists~\cite{Baumgartner_2008, Rivas2012, Evans1977}.
    Furthermore, since $\mathscr{L} [\hat{r}_k^\dag] = (\mathscr{L} [\hat{r}_k])^\dag$, if $\hat{r}_k$ is an eigenmode with eigenvalue $\Lambda_k$, its conjugate $\hat{r}_k^\dag=\hat{r}_{k'}$ is also an eigenmode, with eigenvalue $\Lambda_{k'}=\Lambda_k^*$.

    The Liouvillian eigenspectrum captures the dynamical characteristics of the system. 
    The negative real part of the Liouvillian eigenvalue, $\Gamma_k \equiv -\operatorname{Re} [\Lambda_k]$, determines the decay rate of the $k$-th eigenmode. Depending on whether the decay rate is zero, the eigenmodes can be categorized into two groups: modes with zero decay rates are undamped, whereas modes with non-zero decay rates are damped.
    Because of their non-zero decay rates, damped modes gradually decay and eventually vanish, leaving only undamped modes in the long-time dynamics.
    The imaginary part of the Liouvillian eigenvalue, $\omega_k \equiv \operatorname{Im}[\Lambda_k]$, corresponds to the oscillation frequency. This allows the eigenmodes to be classified into four types: persistent oscillatory, steady, underdamped, and overdamped modes.
    A comprehensive classification of the eigenmodes is presented in Table~\ref{table: eigenmodes_classification}. 
        
    \begin{table}[hbtp]
        \caption{The classification of Liouvillian eigenmodes.}
        \label{table: eigenmodes_classification}
        \begin{ruledtabular}
            \begin{tabular}{c|cc}
                & 
                $\Gamma_k \equiv - \operatorname{Re} [\Lambda_k] = 0$ \footnotemark[1] &
                $\Gamma_k \equiv - \operatorname{Re} [\Lambda_k] > 0$ \footnotemark[2] \\
                \colrule
                $\omega_k \equiv \operatorname{Im} [\Lambda_k] \neq 0$ & Persistent oscillatory & Underdamped \\
                $\omega_k \equiv \operatorname{Im} [\Lambda_k] = 0$ & Steady & Overdamped \\
            \end{tabular}
            \begin{minipage}[t]{0.45\linewidth}
                \footnotetext[1]{Undamped modes}
            \end{minipage}
            \hfill
            \begin{minipage}[t]{0.45\linewidth}
                \footnotetext[2]{Damped modes}
            \end{minipage}
        \end{ruledtabular}
    \end{table}

    The persistent oscillatory eigenmodes are of particular interest among the four types of eigenmodes. Instead of reaching a stationary phase in the long-time limit, the observable of the open quantum system exhibits periodic behavior. However, such eigenmodes are typically absent in conventional quantum systems because of inevitable dissipation. In the following, we develop a mathematical framework for engineering open quantum systems that support persistent oscillatory modes.

% \clearpage
\section{Mathematical framework}

    Before presenting our method for constructing persistent oscillatory modes, we first review the decoherence-free framework and the conditions that allow a system to exhibit persistent oscillatory modes in the following section. This is a widely used approach for engineering states in open quantum systems. We then extend these conditions to present our novel generalized result.

\subsection{\label{sec: DFS} Decoherence-free Subspaces}

   We assume that there exists a subspace $\mathcal{W} \subseteq \mathcal{H}$ in Hilbert space $\mathcal{H}$, which satisfies the following conditions,
    \begin{enumerate}
        \item $\mathcal{W}$ is invariant under $\hat{H}$, each $\hat{L}_i$, and each $\hat{L}_i^\dagger$,i.e., 
        \begin{equation*}
            \hat{H}(\mathcal{W}) \subseteq \mathcal{W}, \quad 
            \hat{L}_i(\mathcal{W}) \subseteq \mathcal{W}, \quad 
            \hat{L}_i^\dagger(\mathcal{W}) \subseteq \mathcal{W}, 
            \quad \forall i,
        \end{equation*}
        \item The restriction of each jump operator $\hat{L}_i$ to $\mathcal{W}$ is proportional to the identity operator, i.e. 
        \begin{equation*}
            \hat{L}_i|_{\mathcal{W}} = \xi_i \, \mathbb{I}_{\mathcal{W}}, 
            \quad \xi_i \in \mathbb{C}, \ \forall i.
        \end{equation*}
        Equivalently, this can be expressed using state $\ket{w}$ that belongs to the subspace $\mathcal{W}$, 
        \begin{equation*}
            \hat{L}_i\ket{w} = \xi_{i} \ket{w}, \ \forall \ket{w} \in \mathcal{W} .
        \end{equation*}
    \end{enumerate}
    Then, $\mathcal{W}$ is a decoherence-free subspace, and the Liouvillian superoperator $\mathscr{L}$ has multiple, undamped modes.
    
    To prove the above statement, it is convenient to work in the eigenbasis $\mathcal{E}_H$ of Hamiltonian $\hat{H}$. In this basis, each jump operator $\hat{L}_i$ that satisfies the specified conditions can be represented by
    \begin{equation}
        \hat{L}_i = \Big( \xi_{i} \sum_{\substack{\ket{f} \in \\ \mathcal{E}_H \cap \mathcal{W}}} \ket{f} \bra{f} \Big) + \sum_{\substack{\ket{p}, \ket{q} \in \\ \mathcal{E}_H \cap \mathcal{W}^\perp}} \Big( \mu_{i}^{pq} \, \ket{p} \bra{q} \Big) , 
    \label{eq: jump operator expression}
    \end{equation}
    where $\xi_{i}, \mu_{i}^{pq} \in \mathbb{C}$ and $\mathcal{W}^\perp$ is the orthogonal complement of $\mathcal{W}$.
    In other words, the matrix representation of the jump operators $\hat{L}_i$ in $\mathcal{E}_H$ is,
    \begin{equation}
        \hat{L}_i = 
            \begin{pmatrix}
                \xi_i & 0 & 0 & 0 & 0 & 0 \\
                0 & \ddots & 0 & 0 & 0 & 0 \\
                0 & 0 & \xi_i & 0 & 0 & 0 \\
                0 & 0 & 0 & \mu_i^{11} & \mu_i^{12} & \cdots \\
                0 & 0 & 0 & \mu_i^{21} & \mu_i^{22} & \cdots \\
                0 & 0 & 0 & \vdots & \vdots & \ddots
            \end{pmatrix}
            _{\mathcal{E}_H}.
    \label{eq: fragmentation}
    \end{equation}
    Consequently, it can be readily shown that
    \begin{equation}
        \hat{r}_{u,v} \equiv \ket{u} \bra{v} , 
    \label{eq: eigenmode expression}
    \end{equation}
    where $\ket{u}, \ket{v} \in \mathcal{E}_H \cap \mathcal{W}$, renders all dissipators zero, 
    \begin{equation}
        \mathcal{D}_i [\hat{r}_{u,v}] = 0, \quad \forall i. 
        \label{eq:dissipaor}
    \end{equation}
    Moreover, $\hat{r}_{u,v}$ are the eigenmodes of the Liouvillian superoperator $\mathscr{L}$, with eigenvalues $\Lambda_{u,v} = -\mathbbm{i} (\lambda_u - \lambda_v)$, where $\lambda_u$ and $\lambda_v$ $\in \mathbb{R}$ are the Hamiltonian eigenvalues corresponding to the states $\ket{u}$ and $\ket{v}$. In other words, we obtain the following result,
    \begin{equation}
        \mathscr{L} [\hat{r}_{u,v}] = -\mathbbm{i} (\lambda_u - \lambda_v) \, \hat{r}_{u,v}. 
    \label{eq: liouvillian eigenvalue}
    \end{equation}

    Because the Liouvillian eigenvalues $\Lambda_{u,v}$ are purely imaginary, the corresponding eigenmodes $\hat{r}_{u,v}$ represent undamped modes. When $\lambda_u \neq \lambda_v$, these modes correspond to persistent oscillatory modes, with frequencies given by the energy gap between the associated Hamiltonian eigenstates. In contrast, when $\lambda_u = \lambda_v$, steady modes are obtained, such as $\hat{r}_{u,u}$ and $\hat{r}_{v,v}$. In both cases, the mode is immune to decay.

%\clearpage
\subsection{\label{sec: general condition} Generalized framework}

    In this section, we develop a generalized condition for the existence of persistent oscillatory modes. In contrast to the decoherence-free subspace approach, the dissipator Eq.~\eqref{eq:dissipaor} are generally non-zero in our framework. Under such general conditions, our approach naturally recovers the decoherence-free subspace method, which becomes a special case of the general framework.

    Our insight begins by promoting the matrix elements of the jump operators $\hat{L}_i$ in Eq.~\eqref{eq: fragmentation} from scalars to the square matrices. In an appropriate orthonormal basis $\mathcal{A}$, the jump operators $\hat{L}_i$ and the Hamiltonian $\hat{H}$ can then be expressed in block-diagonal form as,
    \begin{equation}
        \hat{L}_i = 
            \begin{pmatrix}
                {\bf \Xi_i} & {\bf 0} & {\bf 0} \\
                {\bf 0} & {\bf \Xi_i} & {\bf 0} \\
                {\bf 0} & {\bf 0} & {\bf M_i}
            \end{pmatrix}
            _\mathcal{A}
        ,
        \hat{H} = 
            \begin{pmatrix}
                {\bf H^a} & {\bf 0} & {\bf 0} \\
                {\bf 0} & {\bf H^b} & {\bf 0} \\
                {\bf 0} & {\bf 0} & {\bf H^\text{res}}
            \end{pmatrix}
            _\mathcal{A}
        .
    \label{eq: fragmentation_general}
    \end{equation}
    Here, ${\bf \Xi_{i}}$, $\bf H^a$, $\bf H^b$ are $n \times n$ matrices, while ${\bf M_i}$ and $\bf H^\text{res}$ are $m \times m$ matrices. The matrices ${\bf \Xi_{i}}$ and ${\bf M_i}$ are arbitrary, with no additional constraints imposed, while $\bf H^a$, $\bf H^b$, and $\bf H^\text{res}$ are Hermitian as inherited from $\hat{H}$.
    Besides, $\bf 0$ denote zero matrices
    \footnote{For simplicity, the dimensions of the zero matrices are not explicitly indicated, as they vary throughout the paper. However, their sizes can be inferred from the surrounding context.}. 
    
    To streamline the discussion, the ${\bf M_i}$ and $\bf H^\text{res}$ blocks are omitted in this section without loss of generality. In this simplified setting, the Liouvillian superoperator $\mathscr{L}$ comprises a Hamiltonian $\hat{H}$ and jump operators $\hat{L}_i$ with matrix representations,
    \begin{equation}
        \hat{H} = 
            \begin{pmatrix}
                {\bf H^a} & {\bf 0} \\
                {\bf 0} & {\bf H^b}
            \end{pmatrix}
            _\mathcal{A} 
        ,
        \hat{L}_i = 
            \begin{pmatrix}
                {\bf \Xi_i} & {\bf 0} \\
                {\bf 0} & {\bf \Xi_i}
            \end{pmatrix}
            _\mathcal{A}
        .
    \label{eq: fragmentation_general_chop}
    \end{equation}
    To find the persistent oscillatory mode, we make an ansatz for the eigenmode,
    \begin{equation}
        \hat{r} = 
            \begin{pmatrix}
                {\bf 0} & {\bf 0} \\
                {\bf R} & {\bf 0}
            \end{pmatrix}
            _\mathcal{A} 
        ,
    \end{equation}
    where $\bf R$ is an $n \times n$ matrix. 
    Then, the Liouvillian superoperator acting on the eigenmode $\hat{r}$ yields,
    \begin{equation}
        \mathscr{L} [\hat{r}] = 
            \begin{pmatrix}
                {\bf 0} & {\bf 0} \\
                {\bf Q} & {\bf 0}
            \end{pmatrix}
            _\mathcal{A} 
        ,
    \end{equation}
    where the matrix $\bf Q$ reads,
    \begin{subequations}
    \begin{align}
        \bf Q
        =& -\mathbbm{i} (\bf H^b R - R H^a) \notag\\ 
        &+ \sum_i \gamma_i ( \bf \Xi_i R \Xi^\dag_i - \frac{1}{2} \Xi^\dag_i \Xi_i R - \frac{1}{2} R \Xi^\dag_i \Xi_i) \\
        =& \ \mathbbm{i} {\bf \Delta_H R } + \mathscr{L}^{\sharp} [{\bf R}].
    \end{align}
    \end{subequations}
    Here, $\bf \Delta_H \equiv H^a - H^b$, and
    \begin{subequations}
    \begin{align}
        \mathscr{L}^{\sharp} [{\bf R] } 
        \equiv & -\mathbbm{i} [{\bf H^a, R }] + \sum_i \gamma_i \, \mathcal{D}^\sharp_i [\bf R] ,\\
        \mathcal{D}^\sharp_i [\bf R] 
        \equiv & \ \bf \Xi_i R \Xi^\dag_i - \frac{1}{2} \Xi^\dag_i \Xi_i R - \frac{1}{2} R \Xi^\dag_i \Xi_i .
    \end{align}
    \end{subequations}

    Since $\mathscr{L}^{\sharp}$ is also of Lindbladian form, it is guaranteed to have a steady mode $\bf R_*$ such that 
    \begin{equation}
        \mathscr{L}^{\sharp} [{\bf R_* }] = 0 . 
    \end{equation}
    Consequently, if $\bf R = R_*$ and $\bf R_*$ acts as an eigenmode of $\bf \Delta_H$, referred to as the ``weak $\bf \Delta_H$ condition'', namely,
    \begin{equation}
        {\bf \Delta_H R_* } = \omega {\bf R_* }
    \label{eq: weak Delta H condition}
    \end{equation}
    for some non-zero real scalar $\omega$, the corresponding eigenmode $\hat{r}$ is a persistent oscillatory mode with eigenvalue $+ \mathbbm{i} \omega$, i.e.,
    \begin{equation}
        \mathscr{L} [\hat{r}] = + \mathbbm{i} \omega \, \hat{r} .
    \end{equation}
    In particular, a stronger sufficient condition is  
    \begin{equation}
        {\bf \Delta_H } = \omega \, \mathbb{I}_n , 
    \label{eq: strong Delta H condition}
    \end{equation}
    where $\mathbb{I}_{n}$ is an $n$-dimensional identity operator, which is referred to as the ``strong $\bf \Delta_H$ condition''.
    Note that $\omega$ determines the oscillation frequency of the persistent oscillatory eigenmode $\hat{r}$. 
    When $\omega = 0$, the  oscillatory mode regresses to a steady mode. 
    
    A key distinction between the strong and weak $\bf \Delta_H$ condition concerns the robustness of the resulting oscillatory modes. Oscillatory modes whose existence relies on the strong condition are robust against perturbations, whereas those arising under the weak condition are sensitive to perturbations. This will be illustrated in Sec.~\ref{sec: Example 3} through an artificial but instructive example.

    As an additional remark, $\bf R_*$ is generally not unique, and multiple solutions may exist. Nevertheless, at least one of these is Hermitian~\cite{Baumgartner_2008, Rivas2012, Evans1977}.
    For this particular Hermitian $\bf R_*$, we can find a conjugate persistent oscillatory mode. Its block diagonal form using the same basis reads,
    \begin{equation}
            \begin{pmatrix}
                {\bf 0} & {\bf R_*} \\
                {\bf 0} & {\bf 0}
            \end{pmatrix}
            _\mathcal{A}.
    \end{equation}
  At the same time, 
    \begin{equation}\label{steady_state_A}
        \begin{pmatrix}
            {\bf R_*} & {\bf 0} \\
            {\bf 0} & {\bf 0}
        \end{pmatrix}
        _\mathcal{A}  
        \text{ and }
        \begin{pmatrix}
            {\bf 0} & {\bf 0} \\
            {\bf 0} & {\bf R_*}
        \end{pmatrix}
        _\mathcal{A} 
    \end{equation} 
    represent the steady modes of the Liouvillian $\mathscr{L}$.
   
    In summary, a sufficient condition for the emergence of persistent oscillatory eigenmodes is that the Hamiltonian $\hat{H}$ and the jump operators $\hat{L}_i$ take the forms specified in Eq.~\eqref{eq: fragmentation_general}, together with the holding of the $\Delta H$ condition (either in its weak form Eq.~\eqref{eq: weak Delta H condition} or or in its strong form Eq.~\eqref{eq: strong Delta H condition}).
    Importantly, the construction of the persistent oscillatory mode requires $\mathscr{L}^\sharp [{\bf R_* }] = 0$ but not $\mathcal{D}^\sharp [\bf R_*]$, meaning that the dissipator $\mathcal{D} [\hat{r}]$ is non-vanishing in general. The discussion above shows that the decoherence-free subspace method emerges as a special case, i.e. when the dimension $n$ is taken to be 1.

% \clearpage
\section{\label{sec: Physical Interpretation} Physical Interpretation of the $\bf \Delta_H$ conditions}

    It is natural to ask what physical mechanism underlies the role of the $\bf \Delta_H$ condition as a sufficient condition for the emergence of persistent oscillatory eigenmodes. The answer lies in the separability of the system ensured by the $\bf \Delta_H$ condition.

    A closed system is said to be separable (into two subsystems $\mathcal{S}_{\mathrm{coh}}$ and $\mathcal{S}_{\mathrm{dec}}$) if its Hamiltonian can be decomposed as 
    \begin{equation}
        \hat{H} = ( \mathbb{I}^{(\mathrm{coh})} \otimes \hat{H}^{(\mathrm{dec})} ) + ( \hat{H}^{(\mathrm{coh})} \otimes \mathbb{I}^{(\mathrm{dec})}) ,
    \label{eq: strong separable H}
    \end{equation}
    for some subsystem Hamiltonians $\hat{H}^{(\mathrm{coh})}$ and $\hat{H}^{(\mathrm{dec})}$. Here $\hat{H}^{(\mathrm{coh})}$ and $\hat{H}^{(\mathrm{dec})}$ act only on subsystems $\mathcal{S}_{\mathrm{coh}}$ and $\mathcal{S}_{\mathrm{dec}}$, respectively, while $\mathbb{I}^{(\mathrm{coh})}$ and $\mathbb{I}^{(\mathrm{dec})}$ are the corresponding identity operators acting on subsystems $\mathcal{S}_{\mathrm{coh}}$ and $\mathcal{S}_{\mathrm{dec}}$, respectively.
    For an open system, separability further requires that the jump operators, in addition to the Hamiltonian, admit the same subsystem decomposition. Namely, the jump operators take the form
    \begin{equation}
        \hat{L}_i = ( \mathbb{I}^{(\mathrm{coh})} \otimes \hat{L}_i^{(\mathrm{dec})} ) + ( \hat{L}_i^{(\mathrm{coh})} \otimes \mathbb{I}^{(\mathrm{dec})}) ,
    \label{eq: strong separable L}
    \end{equation}
    where $\hat{L}_i^{(\mathrm{coh})}$ and $\hat{L}_i^{(\mathrm{dec})}$ act only on subsystems $\mathcal{S}_{\mathrm{coh}}$ and $\mathcal{S}_{\mathrm{dec}}$, respectively.

    When the Hamiltonian $\hat{H}$ and jump operators $\hat{L}_i$ exhibit the block-diagonal structure given in Eq.\eqref{eq: fragmentation_general_chop}, they can be rewritten as
    \begin{subequations}
    \begin{align}
        \hat{H} 
        =& \ ( \mathbb{I}_2 \otimes \frac{1}{2} {\bf H'} ) + ( \sigma^z \otimes \frac{1}{2} {\bf \Delta_H}) , 
        \label{eq: block subsystem H} \\
        \hat{L}_i 
        =& \ \phantom{(} \mathbb{I}_2 \otimes {\bf \Xi_i} ,
        \label{eq: block subsystem L} 
    \end{align}
    \label{eq: block subsystem} 
    \end{subequations}
    where $\bf H' \equiv H^a + H^b$, $\bf \Delta_H \equiv H^a - H^b$, and $\sigma^z$ denotes the Pauli-Z matrix. From this representation, the full system can be viewed as a composition a $2$-dimensional subsystem $\mathcal{S}_{\mathrm{coh}}$ and an $n$-dimensional subsystem $\mathcal{S}_{\mathrm{dec}}$. 
    It is evident that the Hamiltonian $\hat{H}$ is not separable because of the presence of the $\mathbf{\Delta_H}$ term, whereas the jump operators are separable. 
    Physically, the form of jump operators describes a situation in which only subsystem $\mathcal{S}_{\mathrm{dec}}$ is directly interacts with the dissipative environment, while the subsystem $\mathcal{S}_{\mathrm{coh}}$ is not.
    However, this does not imply that subsystem $\mathcal{S}_{\mathrm{coh}}$ is unaffected by the dissipative environment. The non-separability of the Hamiltonian couples the two subsystems, allowing the dissipative dynamics acting on subsystem $\mathcal{S}_{\mathrm{dec}}$ to influence subsystem $\mathcal{S}_{\mathrm{coh}}$ indirectly.

    If the Hamiltonian $\hat{H}$ is also separable, then the full system is genuinely separable, and there is no coupling between the two subsystems. Consequently, the dissipative dynamics acting on subsystem $\mathcal{S}_{\mathrm{dec}}$ cannot propagate to subsystem $\mathcal{S}_{\mathrm{coh}}$, leaving subsystem $\mathcal{S}_{\mathrm{coh}}$ genuinely unaffected by the dissipative environment. As a result, the coherent dynamics of subsystem $\mathcal{S}_{\mathrm{coh}}$ is preserved and manifest as persistent oscillatory behavior of the full system. It will be shown below that the $\mathbf{\Delta}_H$ conditions imply the separability and thereby leads to the existence of persistent oscillatory modes.

    It is straightforward to verify that the strong $\mathbf{\Delta}_H$ condition, Eq.~\eqref{eq: strong Delta H condition}, ensures the separability of the Hamiltonian by directly substituting $\mathbf{\Delta}_H = \omega \, \mathbb{I}_n$ into Eq.~\eqref{eq: block subsystem H}. 
    On the other hand, establishing the separability of Hamiltonian under the weak $\mathbf{\Delta}_H$ condition, Eq.~\eqref{eq: weak Delta H condition}, is more subtle. 
    The separability is established by the restricting the Hamiltonian $\hat{H}$ to a subspace $U = \{ \tilde{r} \mid \tilde{r} = {\bf K \otimes R_*} , \ {\bf K} \in \mathbb{C}^{2 \times 2} \}$, i.e. 
    \begin{subequations}
    \begin{align}
        \hat{H} \tilde{r}
        =& \ \left( ( \mathbb{I}_2 \otimes \frac{1}{2} {\bf H'} ) + ( \sigma^z \otimes \frac{1}{2} {\bf \Delta_H}) \right ) \tilde{r} \\
        =& \ ( \mathbb{I}_2 \otimes \frac{1}{2} {\bf H'} ) \tilde{r} + ( \sigma^z {\bf K} \otimes \frac{1}{2} {\bf \Delta_H} {\bf R_*} ) \\
        =& \ ( \mathbb{I}_2 \otimes \frac{1}{2} {\bf H'} ) \tilde{r} + ( \sigma^z {\bf K} \otimes \frac{1}{2} \omega {\bf R_*} ) \\
        =& \ \left( ( \mathbb{I}_2 \otimes \frac{1}{2} {\bf H'} ) + ( \sigma^z \otimes \frac{1}{2} \omega \, \mathbb{I}_n ) \right) \tilde{r} ,
    \end{align}
    \end{subequations}
    for all $\tilde{r}$ in subspace $U$. 
    Therefore, the modes within the subspace $U$ are decoupled from the dissipative dynamics and manifest persistent oscillatory behavior in the full system.

% \clearpage
\subsection{\label{sec: Discussion} Discussion}

    This structure of decoupled subsystems (particularly Eq.~\eqref{eq: block subsystem L}) is reminiscent of the concept of decoherence-free subsystems~\cite{PhysRevA.63.042307, Lidar2003, dutta2025introductionmarkovianopenquantum}. However, our result is not merely a special case of the decoherence-free subsystem framework. Rather, our formulation developed here, especially the condition on $\bf \Delta_H$ given by Eq.~\eqref{eq: weak Delta H condition} or \eqref{eq: strong Delta H condition}, is complementary to the existing theory. 

    The common intuition is that a subsystem remains insensitive to the jump operators $\hat{L}_i$ because of the identity operator $\mathbb{I}$, and therefore remains decoherence-free, allowing the full system to support coherent dynamics. However, this perspective does not always provide a complete description and is insufficient for establishing the separability discussed above. In general, the Hamiltonian $\hat{H}$ and the jump operators $\hat{L}_i$ play equally crucial roles in determining the existence of undamped (persistent oscillatory) dynamics.

%\clearpage
\section{\label{sec: Examples} Examples}
    
    In the following sections, we present four examples of model systems that display persistent oscillatory modes. The first two examples can be understood within the framework of decoherence-free subspaces. The latter two are closely related to our framework. A key difference is that the dissipators associated with these persistent oscillatory modes are non-zero, whereas the decoherence-free subspace framework normally requires their disappearance.
    
\subsection{\label{sec: Example 0} Dephasing Quantum Harmonic Oscillator}
    
\begin{figure}[htbp]
        \centering
        \includegraphics[width=1.0\columnwidth]{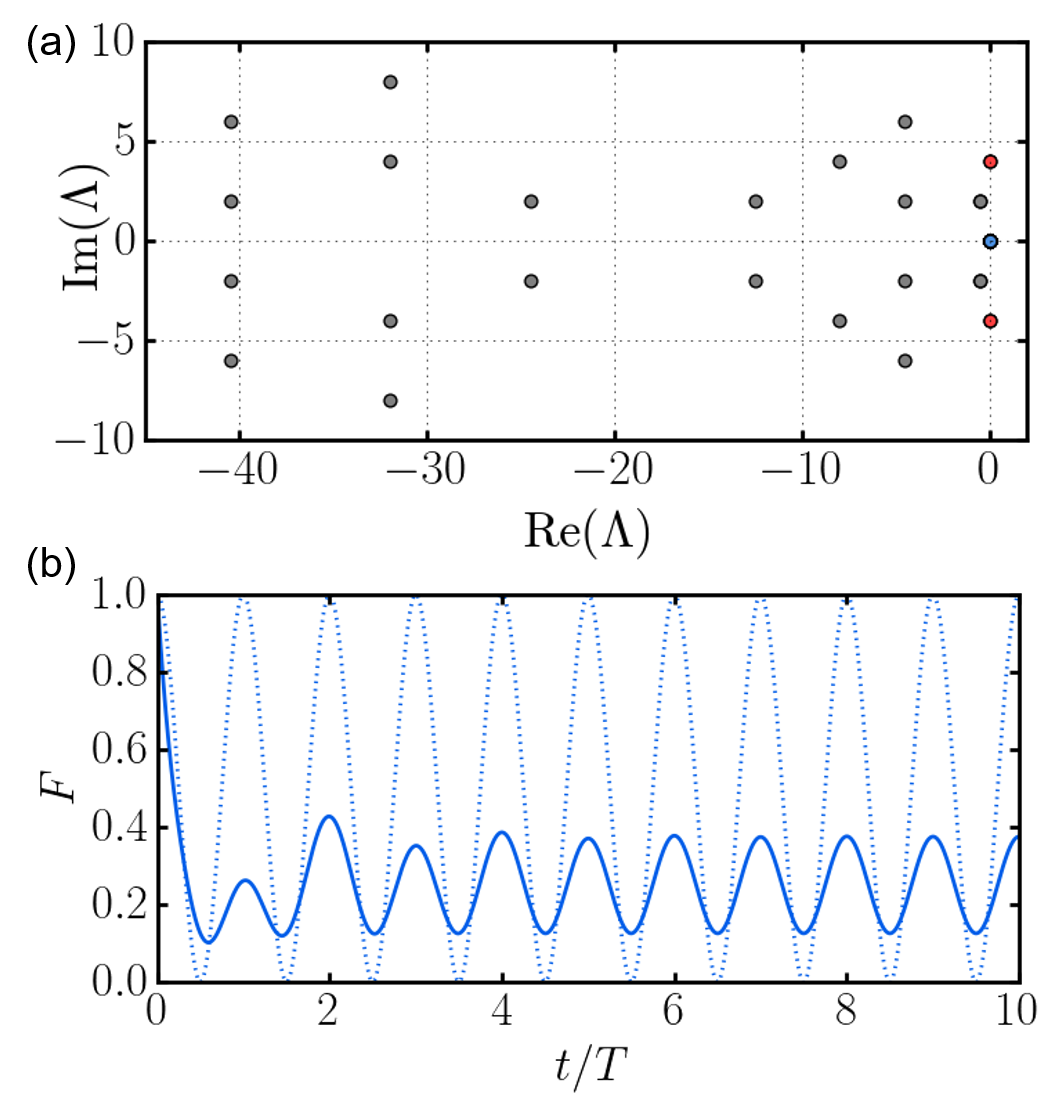}
        \caption{
        (a) Liouvillian eigenspectrum. We only show the spectrum near the steady state. The red, blue, and gray points indicate oscillatory, steady, and damped modes, respectively. Notably, the spectrum features an eigenvalue pair at $\pm \mathbbm{i} 4$.
        (b) Time evolution of fidelities. The dashed curve shows the fidelity between $\rho_\text{osc}(t)$ and $\rho_\text{osc}(0)$. Solid curve shows the fidelity between $\rho_\text{damp}(t)$ and $\rho_\text{damp}(0)$.
        The parameters $\{\nu, \gamma\} = \{2, 1\}$ are used throughout for this Dephasing Quantum Harmonic Oscillator Example, yielding $T = 0.5\pi$. 
        }
        \label{fig: plot_0}
    \end{figure}

    We begin with a simple example to demonstrate how to construct persistent oscillatory modes using the decoherence-free subspace method. We consider a harmonic oscillator with the Hamiltonian ($\hbar\equiv 1$) and jump operator given by,
    \begin{equation}
        \hat{H} = \nu \,\hat{N} \text{\quad and \quad}
        \hat{L} =  (\hat{N} - 1)^2 ,
    \end{equation}
    where $\hat{N}=\hat{a}^{\dagger}\hat{a}$ is the number operator with $\hat{a}^{\dagger}$ ($\hat{a}$) to be the bosonic creation (annihilation) operator. The oscillator frequency is $\nu$. 
    Since the Fock (number) states $\ket{N}$ are eigenstates of both the Hamiltonian and the jump operator, any subspace formed from these states remains invariant under $\hat{H}$, $\hat{L}$, and $L^\dag$. 
    Among the Fock states, states $\ket{0}$ and $\ket{2}$ are particularly interesting as,
    \begin{equation}
        \hat{L} \ket{0} = \ket{0} \text{\quad and \quad} \hat{L} \ket{2} = \ket{2} ,
    \end{equation}
   This indicates that they form a subspace $\mathcal{W} = \operatorname{span} [\{ \ket{0}, \ket{2} \}]$, which is the decoherence-free subspace of the system.
    Quantum states described by operators 
    \begin{equation}
        \hat{r}_{0,2} = \ket{0} \bra{2} \text{ and } \hat{r}_{2,0} = \ket{2} \bra{0} 
    \end{equation}
    correspond to the persistent oscillatory modes, with Liouvillian eigenvalues $\Lambda = \mp \mathbbm{i} 2 \nu$ and oscillation frequencies $\omega = \mp 2 \nu$. One can also find that 
    \begin{equation}
        \hat{r}_{0,0} = \ket{0} \bra{0} \text{ and } \hat{r}_{2,2} = \ket{2} \bra{2} 
    \end{equation}
    are the steady modes. The Liouvillian eigenspectrum shown in Fig.~\ref{fig: plot_0}~(a) features two purely imaginary Liouvillian eigenvalues that give rise to the two anticipated persistent oscillatory modes. 

    We now show that the oscillatory modes can be excited in the dissipative dynamics. The system is initialized in a nontrivial superposition of the two states so that its density matrix consists only of oscillatory and steady modes. For example, when the initial state is,
    \begin{equation}
         \ket{s_\text{osc}(0)} = \frac{1}{\sqrt{2}} (\ket{0} + \ket{2}) , 
    \end{equation}
    the initial density matrix reads,
    \begin{equation}
        \rho_\text{osc} (0) = \frac{1}{2} \Big( \ket{0} \bra{2} + \ket{2} \bra{0} + \ket{0} \bra{0} + \ket{2} \bra{2} \Big) ,
    \end{equation}
    which evolves as
    \begin{equation}
    \begin{split}
        \rho_\text{osc} (t) = 
        & \frac{1}{2} \Big( e^{\mathbbm{i} 2 \nu t} \, \ket{0} \bra{2} + e^{- \mathbbm{i} 2 \nu t}\, \ket{2} \bra{0} \\
        & + \ket{0} \bra{0} + \ket{2}\bra{2} \Big)
    \end{split}
    \end{equation}
    The dashed curve in Fig.~\ref{fig: plot_0}~(b) 
    shows the time evolution of the fidelity $F[\rho_1, \rho_2] \equiv \left(\operatorname{Tr} \sqrt{\sqrt{\rho_1} \rho_2 \sqrt{\rho_1}}\right)^2$~\cite{Jozsa01121994} between the  density matrix $\rho_\text{osc}(t)$ and its initial state $\rho_\text{osc}(0)$. The fidelity oscillates at a period $T = \frac{\pi}{\nu}$. The complete revival confirms that the oscillation is persistent.

    Alternatively, if the initial state does not lie fully within the decoherence-free subspace, e.g.,
    \begin{equation}
         \ket{s_\text{damp}(0)} = \frac{1}{2} (\ket{0} + \ket{1}+ \ket{2} + \ket{3}) , 
    \end{equation}
    The system initially exhibited a combination of damped and oscillatory behavior. During the transient period, the damped modes gradually decay. After this period, the dynamics displays the oscillations characteristic of the decoherence-free subspace, as illustrated by the solid curve in Fig.~\ref{fig: plot_0}~(b).

    As an additional remark, although the dephasing jump operator used in this example has been chosen mainly for the sake of  simplicity, an experimentally feasible dephasing jump operator version can be realized through nonlinear reservoir engineering \cite{rojkov2024stabilizationcatstatemanifoldsusing}, yielding the following jump operator,
    \begin{equation}
        \hat{L} = \sum_{N=0}^\infty \mathcal{G}_N (x) \ket{N}\bra{N} ,
    \end{equation}
   which essentially has the same effect.
    Here, $\mathcal{G}_N(x)$ is the $N$-th order Laguerre polynomial, and $x$ is a tunable parameter. In particular, there exists an $x_*$ such that $\mathcal{G}_0 (x_*) = \mathcal{G}_m (x_*)$ for integer $m > 1$. When $m=2$, this corresponds to the example discussed in this section.

%\clearpage
\subsection{\label{sec: Example 1} Dephasing spin chain}

    In this section, we present an example involving multiple jump operators $\hat{L}_i$ and decoherence-free subspaces $\mathcal{W}_j$. We consider a three-qubit chain subject to local dephasing on the boundary qubits (the first and third), 
    \begin{equation}
        \hat{L}_1 = \hat{\sigma}^z_1 \quad \operatorname{ and } \quad \hat{L}_2 = \hat{\sigma}^z_3 .
    \label{eq: ???}
    \end{equation}
    Although there is no dephasing in the second qubit, it is coupled to the boundary qubits through an interacting Hamiltonian as follows,
    \begin{equation}
        \hat{H} = \chi \hat{\sigma}^z_1 \hat{\sigma}^z_2 \hat{\sigma}^z_3 + J_{12} \hat{\sigma}^z_1 \hat{\sigma}^z_2 + J_{23} \hat{\sigma}^z_2 \hat{\sigma}^z_3 , 
    \label{eq: ???}
    \end{equation}
    where $\chi$ is the three-spin interaction strength and $J_{ij}$ is the two-spin interaction strength between the $i$-th and $j$-th qubits.
    
\begin{figure}[htbp]
        \centering
        \includegraphics[width=0.92\columnwidth]{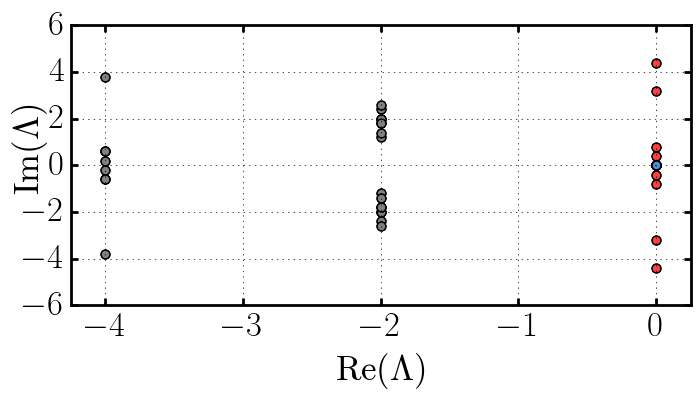}
        \caption{
        Liouvillian eigenspectrum. The red, blue, and gray points indicate the oscillatory, steady, and damped modes, respectively. Notably, the spectrum includes eigenvalue pairs located at $\pm \mathbbm{i} 4.4$, $\pm \mathbbm{i} 3.2$, $\pm \mathbbm{i} 0.8$, and $\pm \mathbbm{i} 0.4$. The parameters $\{ \chi, J_{12}, J_{23}, \gamma_1, \gamma_2 \} = \{ 0.3,0.9,1.0,1.0,1.0 \}$ are used in the calculation.
        }
        \label{fig: plot_1}
    \end{figure}

    For such system, the Hilbert space $\mathcal{H}$ admits four decoherence-free subspaces $\mathcal{W}_j$ $(j=1,2,3,4)$, as listed in Table~\ref{table: general_case_fragmentation}. The restrictions of the two jump operators to each decoherence-free subspace $\hat{L}_i|_{\mathcal{W}_j}$ are equal to either the identity operator or its negative operator. 
    Consequently, each decoherence-free subspace generates two persistent oscillatory modes, yielding a total of eight persistent oscillatory modes, provided that the restriction of the Hamiltonian to each decoherence-free subspace $\hat{H}|_{\mathcal{W}_j}$ is non-degenerate. These modes correspond to $\ket{+_j}\bra{-_j}$ and $\ket{-_j}\bra{+_j}$, where their explicit representations are listed in Table~\ref{table: general_case_fragmentation}. Their eigenvalues are $\pm \mathbbm{i} 2 (\chi+J_{12}+J_{23})$, $\pm \mathbbm{i} 2 (-\chi-J_{12}+J_{23})$, $\pm \mathbbm{i} 2 (-\chi+J_{12}-J_{23})$, and $\pm \mathbbm{i} 2 (\chi-J_{12}-J_{23})$, respectively. 
    In addition, there is an eight-fold degeneracy at zero Liouvillian eigenvalue, signifying the existence of eight steady modes. These modes form disconnected Hilbert subspace in the dissipative regime~\cite{yan2025hilbertspacefragmentationdrivendephasing, PhysRevResearch.5.043239}. To illustrate the persistent oscillatory modes, we provide an example in  Fig.~\ref{fig: plot_1}. The eight persistent oscillatory modes appear in conjugate pairs. 
    % Fig.~\ref{fig: example_1_spectrum}.
    
    %
    \begin{table}[hbtp]
    \centering
    \caption{The fragmentation of the Hilbert space $\mathcal{H}$.}
    \label{table: general_case_fragmentation}
    \begin{ruledtabular}
    \begin{tabular}{ c||c c|c } 
        Subspaces $\mathcal{W}_j$ & $\hat{L}_1|_{\mathcal{W}_j}$ & $\hat{L}_2|_{\mathcal{W}_j}$ & $\hat{H}|_{\mathcal{W}_j}$ \\
        \colrule
        span[$\ket{+_1}, \ket{-_1}$] & $+ \mathbb{I}$ & $+ \mathbb{I}$ & $(\phantom{+} \chi+J_{12}+J_{23}) \, \sigma^z_{\mathcal{W}_1}$ \\ 
        span[$\ket{+_2}, \ket{-_2}$] & $- \mathbb{I}$ & $+ \mathbb{I}$ & $(-\chi-J_{12}+J_{23}) \, \sigma^z_{\mathcal{W}_2}$ \\
        span[$\ket{+_3}, \ket{-_3}$] & $+ \mathbb{I}$ & $- \mathbb{I}$ & $(-\chi+J_{12}-J_{23}) \, \sigma^z_{\mathcal{W}_3}$ \\
        span[$\ket{+_4}, \ket{-_4}$] & $- \mathbb{I}$ & $- \mathbb{I}$ & $(\phantom{+} \chi-J_{12}-J_{23}) \, \sigma^z_{\mathcal{W}_4}$ \\
    \end{tabular}
    \footnotetext[0]{
    $\ket{+_1} = \ket{eee} , \, \ket{-_1} = \ket{ege} ,$
    $\ket{+_2} = \ket{gee} , \, \ket{-_2} = \ket{gge} ,$
    $\ket{+_3} = \ket{eeg} , \, \ket{-_3} = \ket{egg} ,$
    $\ket{+_4} = \ket{geg} , \, \ket{-_4} = \ket{ggg} ,$
    \\
    \noindent $\sigma^z_{\mathcal{W}_j}$ is the Pauli-Z operator in the subspace $\mathcal{W}_j$, $\ket{ijk}$ are the three spins states, with $i,j,k = g,e$, where $g$ and $e$ denote the ground and excited states of the spins, respectively.
    }
    \end{ruledtabular}
    \end{table}
    %

% \clearpage
\subsection{\label{sec: Example 2} Spin chain with collective dissipation}
    
    In this example, we demonstrate the generalized condition for constructing oscillatory modes. This will be achieved using a three-qubit system. The resulting persistent oscillatory modes cannot be explained within the framework of decoherence-free subspaces because their associated dissipators are non-vanishing. As we will show later, in this example $\bf \Xi_i$ is a $2 \times 2$ matrix. 
    
    Let the jump operator of the system be the collective lowering operator,
    \begin{equation}
        \hat{L} = \hat{\sigma}^-_1 + \hat{\sigma}^-_2 + \hat{\sigma}^-_3 ,
    \label{eq: collective lowering operator}
    \end{equation}
    where $\hat{\sigma}^-_i = \ket{\downarrow}_i \bra{\uparrow}_i$ represents the lowering operator acting on the $i$-th qubit. In the uncoupled (computational) basis, \\
    $\mathcal{B} = \{ \ket{\downarrow \downarrow \downarrow}, \ket{\downarrow \downarrow \uparrow}, \ket{\downarrow \uparrow \downarrow}, \ket{\downarrow \uparrow \uparrow}, \ket{\uparrow \downarrow \downarrow}, \ket{\uparrow \downarrow \uparrow}, \ket{\uparrow \uparrow \downarrow}, \ket{\uparrow \uparrow \uparrow}\}$, 
    matrix representation of the jump operator is as follows,
    \begin{equation}
        \hat{L} = 
        \begin{pmatrix}
            0 & 1 & 1 & 0 & 1 & 0 & 0 & 0 \\
            0 & 0 & 0 & 1 & 0 & 1 & 0 & 0 \\
            0 & 0 & 0 & 1 & 0 & 0 & 1 & 0 \\
            0 & 0 & 0 & 0 & 0 & 0 & 0 & 1 \\
            0 & 0 & 0 & 0 & 0 & 1 & 1 & 0 \\
            0 & 0 & 0 & 0 & 0 & 0 & 0 & 1 \\
            0 & 0 & 0 & 0 & 0 & 0 & 0 & 1 \\
            0 & 0 & 0 & 0 & 0 & 0 & 0 & 0 
        \end{pmatrix}
        _\mathcal{B}.
    \end{equation}
    At first glance, it does not exhibit the block-diagonal structure given in Eq. ~\eqref{eq: fragmentation_general}. This can be achieved, however, in the coupled basis, 
    \begin{equation*}
        \mathcal{C} = 
        \left\{
        \begin{array}{l}
            \begin{minipage}{0.45\linewidth}
            $\frac{1}{\sqrt{2}} (\ket{\downarrow \downarrow \uparrow} - \ket{\uparrow \downarrow \downarrow})$, \\
            $\frac{1}{\sqrt{2}} (\ket{\downarrow \uparrow \uparrow} - \ket{\uparrow \uparrow \downarrow })$, 
            \end{minipage}
            \quad
            \begin{tabular}{l}
                doublet
            \end{tabular} \\
            \noalign{\vskip 2pt}
            \noalign{\hrule height 0.4pt}
            \noalign{\vskip 2pt}
            \begin{minipage}{0.45\linewidth}
            $\frac{-1}{\sqrt{6}} (\ket{\downarrow \downarrow \uparrow} + \ket{\uparrow \downarrow \downarrow} - 2 \ket{\downarrow \uparrow \downarrow})$, \\
            $\frac{1}{\sqrt{6}} (\ket{\downarrow \uparrow \uparrow} + \ket{\uparrow \uparrow \downarrow } - 2 \ket{\uparrow \downarrow \uparrow})$,
            \end{minipage}
            \quad
            \begin{tabular}{l}
                doublet
            \end{tabular} \\
            \noalign{\vskip 2pt}
            \noalign{\hrule height 0.4pt}
            \noalign{\vskip 2pt}
            \begin{minipage}{0.45\linewidth}
            $\ket{\downarrow \downarrow \downarrow}$, \\
            $\frac{1}{\sqrt{3}} (\ket{\downarrow \downarrow \uparrow} + \ket{\downarrow \uparrow \downarrow} + \ket{\uparrow \downarrow \downarrow})$, \\
            $\frac{1}{\sqrt{3}} (\ket{\downarrow \uparrow \uparrow} + \ket{\uparrow \downarrow \uparrow} + \ket{\uparrow \uparrow \downarrow })$, \\
            $\ket{\uparrow \uparrow \uparrow}$ 
            \end{minipage}
            \quad
            \begin{tabular}{l}
                quartet
            \end{tabular} 
        \end{array}
        \right\}. 
    \end{equation*}
    The coupled basis is formed by  the computational basis and is orthogonal to each other. As the above equation shows, we have either two doublet sectors, and a quartet sector. 
    Using the coupled basis, the matrix representation of the jump operator reads, 
    \begin{equation}
        \hat{L} =
        \begin{pNiceMatrix}
        0 & 1 & 0 & 0 & 0 & 0 & 0 & 0 \\
        0 & 0 & 0 & 0 & 0 & 0 & 0 & 0 \\
        0 & 0 & 0 & 1 & 0 & 0 & 0 & 0 \\
        0 & 0 & 0 & 0 & 0 & 0 & 0 & 0 \\
        0 & 0 & 0 & 0 & 0 & \sqrt{3} & 0 & 0 \\
        0 & 0 & 0 & 0 & 0 & 0 & 2 & 0 \\
        0 & 0 & 0 & 0 & 0 & 0 & 0 & \sqrt{3} \\
        0 & 0 & 0 & 0 & 0 & 0 & 0 & 0
        \CodeAfter
        % First 2x2 block
        \tikz \draw[thick]
            ([shift={(-2pt,2pt)}]1-1.north west)
            rectangle
            ([shift={(2pt,-2pt)}]2-2.south east);
        % Second 2x2 block
        \tikz \draw[thick]
            ([shift={(-2pt,2pt)}]3-3.north west)
            rectangle
            ([shift={(2pt,-2pt)}]4-4.south east);
        % Last 4x4 block
        \tikz \draw[thick]
            ([shift={(-2pt,3pt)}]5-5.north west)
            rectangle
            ([shift={(6pt,-2pt)}]8-8.south east);
        \end{pNiceMatrix}
        _\mathcal{C} .
    \end{equation}
    It can be observed that this form has the same structure as that of Eq.~\eqref{eq: fragmentation_general}. Explicitly, it can be seen that the matrix  ${
    \bf \Xi } = \begin{psmallmatrix} 0 & 1 \\ 0 & 0 \end{psmallmatrix} $.

    Hamiltonian that can realize 
%    \footnote{???????????????}
    the same block-diagonal structure is the three-body Heisenberg XXX model~\cite{Heisenberg1928, TAKHTAJAN1981231, BAXTER1972193, Yang-Guo-Hui_2007, PhysRevA.74.052105,thakurHeisenbergSpin$frac12$XXZ2018} with  homogeneous external fields, under the open boundary condition, which reads,
    \begin{equation}
    \begin{split}
        \hat{H} =&
            J \left( \sum_{i=1}^{2} \hat{\sigma}_i^x \hat{\sigma}_{i+1}^x 
            + \sum_{i=1}^{2} \hat{\sigma}_i^y \hat{\sigma}_{i+1}^y 
            + \sum_{i=1}^{2} \hat{\sigma}_i^z \hat{\sigma}_{i+1}^z \right) \\
            &+ h_x \sum_{i=1}^{3} \hat{\sigma}_i^x 
            + h_y \sum_{i=1}^{3} \hat{\sigma}_i^y 
            + h_z \sum_{i=1}^{3} \hat{\sigma}_i^z,
    \label{eq: heisenberg hamiltonian}
    \end{split}
    \end{equation}
    $\hat{\sigma}_i^\alpha$ ($\alpha = x,y,z$)  are the Pauli operators acting on the $i$-th qubit,
    $J$ is the coupling constant, and $h_{x/y/z}$ denote the strength of the external fields along the $x$, $y$, and $z$ directions, respectively. To illustrate the block structure, as required by Eq.~\eqref{eq: fragmentation_general}, we give the matrix representation of the Hamiltonian in basis $\mathcal{C}$,
    \begin{subequations}
    \begin{align}
        {\bf H^a } =& 
            \begin{pmatrix}
               h_z & h_x - \mathbbm{i} h_y \\
                h_x + \mathbbm{i} h_y & - h_z
            \end{pmatrix} ,
        \\
         {\bf H^b } =& 
            \begin{pmatrix}
                - 4 J + h_z & h_x - \mathbbm{i} h_y \\
                h_x + \mathbbm{i} h_y & - 4 J - h_z
            \end{pmatrix} ,
        \\
        {\bf H^\text{res}} =&
            \begingroup
            \setlength{\arraycolsep}{0pt}
            \begin{pmatrix}
                4 J + 3 h_z & \sqrt{3} (h_x - \mathbbm{i} h_y) & 0 & 0 \\
                \sqrt{3} (h_x + \mathbbm{i} h_y) & 4 J + h_z & 2 (h_x - \mathbbm{i} h_y) & 0 \\
                0 & 2 (h_x + \mathbbm{i} h_y) & 4 J - h_z & \sqrt{3} (h_x - \mathbbm{i} h_y) \\
                0 & 0 & \sqrt{3} (h_x + \mathbbm{i} h_y) & 4 J - 3 h_z
            \end{pmatrix} 
            \endgroup
        \end{align}
    \end{subequations}
    One can find straightly that ${\bf \Delta_H \equiv H^a - H^b } = 4 J \, \mathbb{I}_2$, satisfying the strong $\bf \Delta_H$ condition in Eq. ~\eqref{eq: strong Delta H condition}. 

\begin{figure}[htbp]
        \centering
        \includegraphics[width=0.92\columnwidth]{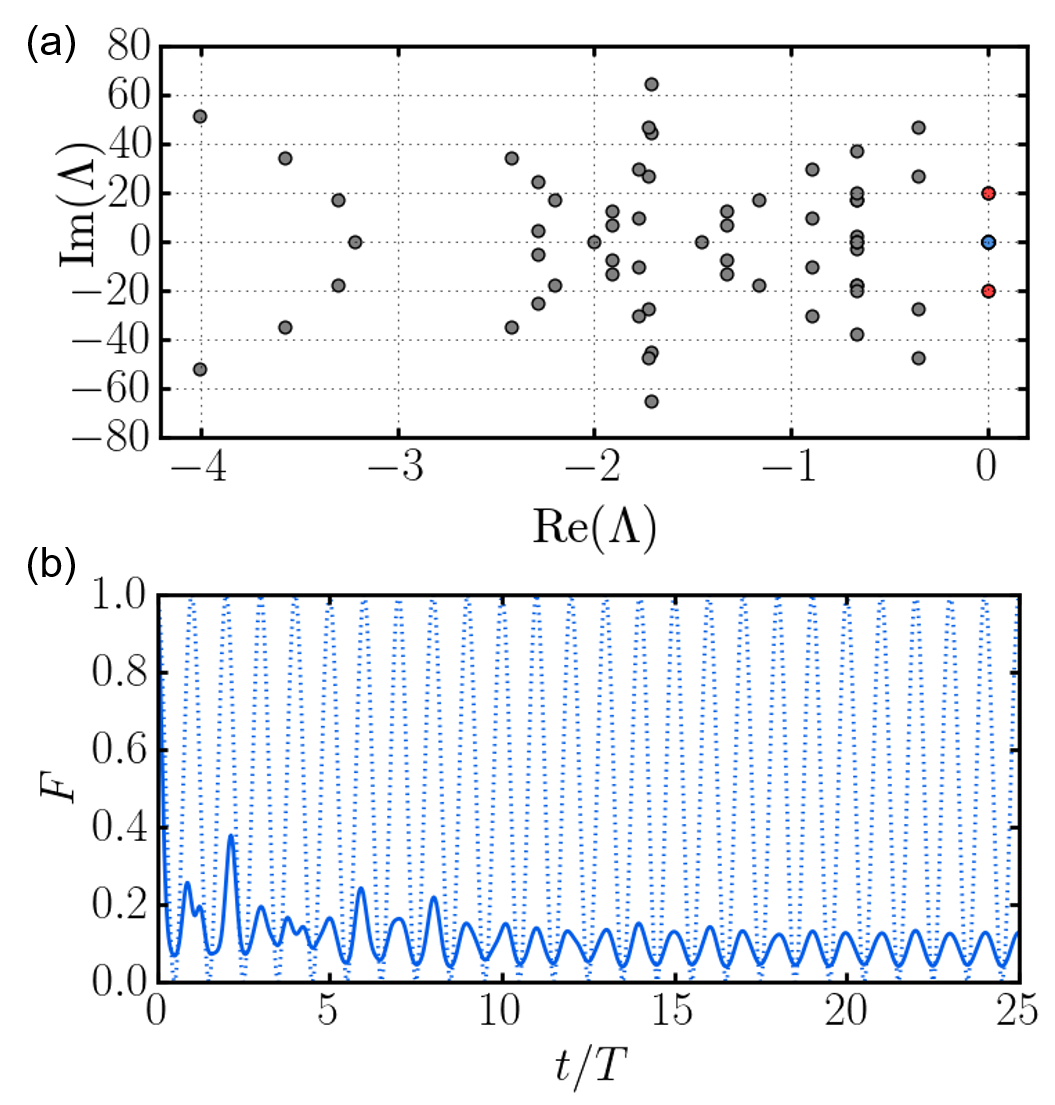}
        \caption{
        (a) Liouvillian eigenspectrum. The red, blue, and gray points indicate the persistent oscillatory, steady, and damped modes, respectively. Notably, the spectrum features an eigenvalue pair at $\pm \mathbbm{i} 20$.
        (b) Time evolution of the fidelities. Dashed curve shows the fidelity between $\rho_\text{osc}(t)$ and $\rho_\text{osc}(0)$. Solid curve shows the fidelity between $\rho_\text{damp}(t)$ and $\rho_\text{damp}(0)$.
        The parameters $\{ J, h_x, h_y, h_z, \gamma \} = \{ 5,5,5,5,1 \}$ are used in the calculation. This yields a period $T = 0.1\pi$ in the oscillation.
        }
        \label{fig: plot_2}
    \end{figure}

    As a result, the Liouvillian $\mathscr{L}$, which comprises the  Hamiltonian in Eq.~\eqref{eq: heisenberg hamiltonian} and the jump operator in Eq.~\eqref{eq: collective lowering operator}, admits a pair of persistent oscillatory eigenmodes. We can explicitly obtain these modes,
    \begin{equation}
        \hat{r} = 
            \begin{pmatrix}
                {\bf 0} & {\bf R_*} & {\bf 0} \\
                {\bf 0} & {\bf 0} & {\bf 0} \\
                {\bf 0} & {\bf 0} & {\bf 0}
            \end{pmatrix}
            _\mathcal{C}
        \text{\quad and \quad}
        \hat{r}^\dag = 
            \begin{pmatrix}
                {\bf 0} & {\bf 0} & {\bf 0} \\
                {\bf R_*} & {\bf 0} & {\bf 0} \\
                {\bf 0} & {\bf 0} & {\bf 0}
            \end{pmatrix}
            _\mathcal{C} ,
    \end{equation}
    where ${\bf R_* } = \frac{8 h_x h_z + 2 h_y \gamma}{8 (h_x^2 + h_y^2 + 2 h_z^2) + \gamma^2} \hat{\sigma}^x + \frac{8 h_y h_z - 2 h_x \gamma}{8 (h_x^2 + h_y^2 + 2 h_z^2) + \gamma^2} \hat{\sigma}^y + \frac{8 h_z^2 + \gamma^2 / 2}{8 (h_x^2 + h_y^2 + 2 h_z^2) + \gamma^2} \hat{\sigma}^z + \frac{1}{2} \mathbb{I}_{2}$. The corresponding eigenvalues are $\Lambda = \mp \mathbbm{i} 4 J$. Importantly, their associated dissipators are non-vanishing, 
    \begin{equation}
        \mathcal{D} [\hat{r}] = 
            \begin{pmatrix}
                {\bf 0} & {\bf D} & {\bf 0} \\
                {\bf 0} & {\bf 0} & {\bf 0} \\
                {\bf 0} & {\bf 0} & {\bf 0}
            \end{pmatrix}
            _\mathcal{C}
        \neq {\bf 0} ,
    \end{equation}
    where ${\bf D} = - \frac{4 h_x h_z + h_y \gamma}{8 (h_x^2 + h_y^2 + 2 h_z^2) + \gamma^2} \hat{\sigma}^x - \frac{4 h_y h_z - h_x \gamma}{8 (h_x^2 + h_y^2 + 2 h_z^2) + \gamma^2} \hat{\sigma}^y + \frac{4 (h_x^2 + h_y^2)}{8 (h_x^2 + h_y^2 + 2 h_z^2) + \gamma^2} \hat{\sigma}^z$, and $\mathcal{D} [\hat{r}^\dag] = (\mathcal{D} [\hat{r}])^\dag \neq {\bf 0}$. It is also worth noting that the existence of these eigenmodes and their oscillation frequencies are independent of the damping rate $\gamma$.
    Fig.~\ref{fig: plot_2}~(a) 
    shows the Liouvillian eigenspectra. The two purely imaginary Liouvillian eigenvalues at $\Lambda = \pm \mathbbm{i} 4J$ are marked with red dots. The corresponding eigenmodes are the oscillatory states.
    
    Next, we examine the oscillatory dynamics. When the system is initialized in a density matrix composed solely of the persistent oscillatory and steady modes, 
    \begin{equation}
        \rho_\text{osc} (0) = \frac{1}{2}
            \begin{pmatrix}
                {\bf R_*} & {\bf R_*} & {\bf 0} \\
                {\bf R_*} & {\bf R_*} & {\bf 0} \\
                {\bf 0} & {\bf 0} & {\bf 0}
            \end{pmatrix}
            _\mathcal{C},
    \end{equation}
    the system oscillates back and forth between the initial state $\rho_\text{osc} (0)$ and state
    \begin{equation}
        \rho_\text{osc} \left(\frac{T}{2}\right) = \frac{1}{2}
            \begin{pmatrix}
                {\bf R_*} & {-\bf R_*} & {\bf 0} \\
                {-\bf R_*} & {\bf R_*} & {\bf 0} \\
                {\bf 0} & {\bf 0} & {\bf 0}
            \end{pmatrix}
            _\mathcal{C} ,
    \end{equation}
    with period $T = \frac{2\pi}{\| \omega \|} = \frac{\pi}{2J}$.
    The dashed curve in Fig.~\ref{fig: plot_2}(b)
    shows a numerical simulation of the fidelity between  the density matrix $\rho_\text{osc}(t)$ and the initial state $\rho_\text{osc}(0)$. 
    This result demonstrates that the system returns to the initial state with perfect fidelity after a period $T$, indicating that the oscillation is persistent and not influenced by the decoherence. 

    If the initial density matrix is also constituted of damped modes, for example,
    \begin{equation}
        \rho_\text{damp} (0) = \ket{\downarrow \downarrow \uparrow} \bra{\downarrow \downarrow \uparrow},
    \end{equation}
    the dynamics does not exhibit complete revival. Once the damped modes have fully decayed, the fidelity oscillates periodically with a lower amplitude.
    % Figure~\ref{fig: example_2_fidelity_damp} 
    The solid curve in Fig.~\ref{fig: plot_2}(b)
    illustrates this case.
    Initially, the fidelity displays irregular dynamics, but subsequently settles into oscillatory behavior.

    Finally, we emphasize that the existence of persistent oscillatory modes is largely independent of the choice of parameters ($J$, $h_x$, $h_y$, $h_z$), and, more importantly, of the damping rate $\gamma$.
    An exception is when $J = 0$, in which case the persistent oscillatory modes regress to steady modes.

\subsection{\label{sec: Example 3} Fine-Tuning Sensitivity: An Example}

    In this artificial but instructive example, a case is demonstrated in which the strong $\bf \Delta_H$ condition is not satisfied, i.e. ${\bf \Delta_H } \neq \omega \, \mathbb{I}_n$. Nevertheless, the parameters can be tuned so that the weak $\bf \Delta_H$ condition, $\bf \Delta_H R_* = \omega R_*$, remains satisfied, thereby generating persistent oscillatory modes. The limitations of this construction, which depend sensitively on the parameter values, are discussed.
    \begin{figure}[hbtp]
    \centering
    \includegraphics[width=0.92\columnwidth]{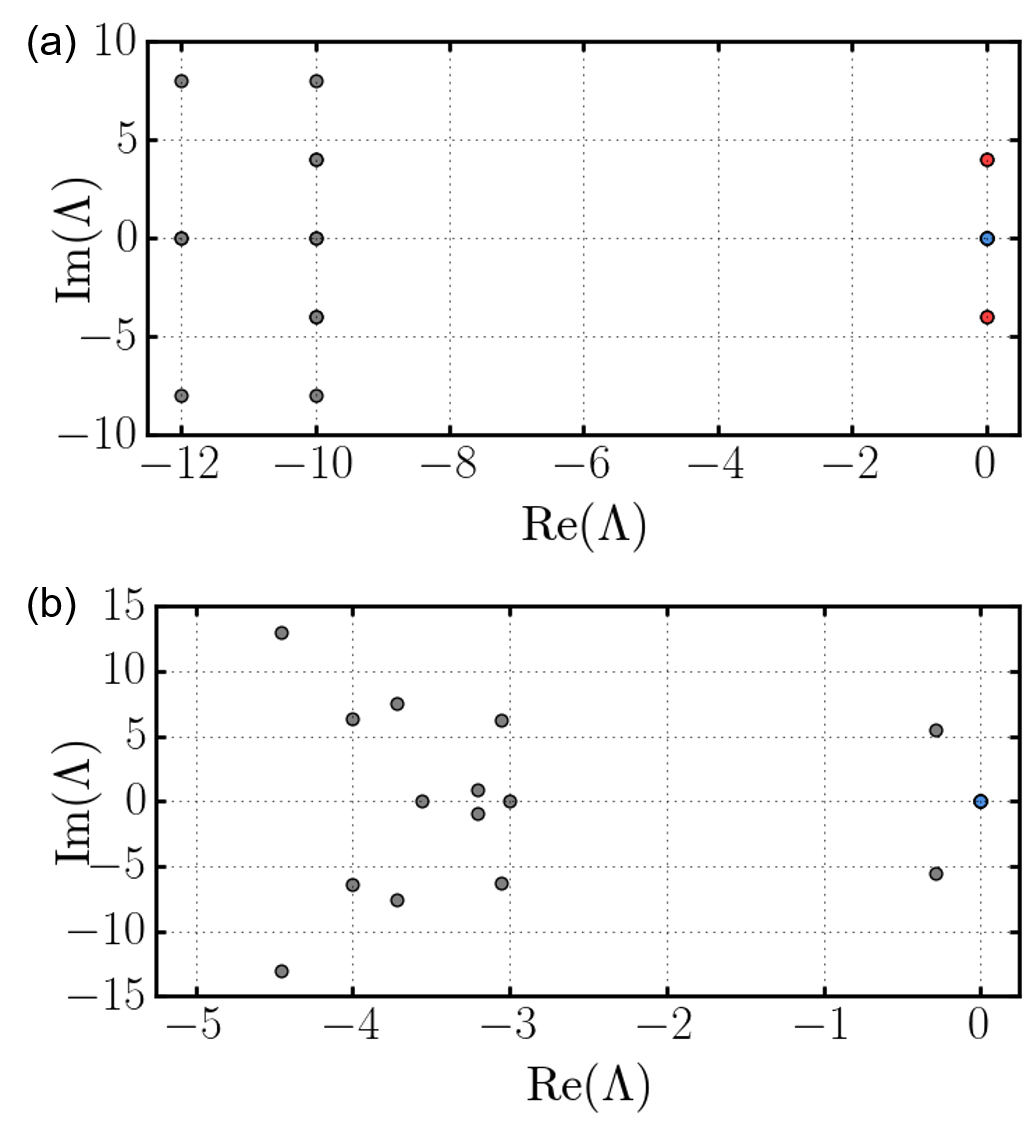}
    \caption{
        Liouvillian eigenspectra. The red, blue, and gray points indicate persistent oscillatory, steady, and damped modes, respectively.
        (a) Parameters $\{E,\gamma_1,\gamma_2\} = \{1,1,8\}$. In this case, the spectrum features an eigenvalue pair at $\pm \mathbbm{i} 4$.
        (b) Parameters $\{E,\gamma_1,\gamma_2\} = \{1,1,1\}$. In this case, the spectrum features no purely imaginary eigenvalues.
    }
    \label{fig: plot_3}
    \end{figure}

    We consider a two-qubit system with a Liouvillian $\mathscr{L}$ characterized by Hamiltonian
        \begin{equation}
            \hat{H} = E \ \Big( 3 \hat{\sigma}_1^z + 6 \hat{\sigma}_2^x + \hat{\sigma}_2^y + \hat{\sigma}_1^z \hat{\sigma}_2^y \Big) ,
        \end{equation}
    and two jump operators given by,
    \begin{equation}
    \begin{split}
        \hat{L}_1 &= \Big( 2 + \mathbbm{i} \hat{\sigma}_2^x - \hat{\sigma}_2^z \Big), \\
        \hat{L}_2 &= \frac{1}{2} \Big( 3 + \mathbbm{i} \hat{\sigma}_2^x + \hat{\sigma}_2^y - \hat{\sigma}_2^z \Big) ,
    \end{split}
    \end{equation}
    where $E$ is the characteristic energy of the system. We will scale the energy with respect to $\gamma_1$. Their matrix representations in the computational basis $\mathcal{B}$ are 
    \begin{equation}
        \hat{H} = 
            E \ \begin{pNiceMatrix}
                3 & 6 - 2\mathbbm{i} & 0 & 0 \\
                6 + 2\mathbbm{i} & 3 & 0 & 0 \\
                0 & 0 & -3 & 6 \\
                0 & 0 & 6 & -3 \\
            \CodeAfter
            % % First 2x2 block
            % \tikz \draw[thick]
            %     ([shift={(-9pt,2pt)}]1-1.north west)
            %     rectangle
            %     ([shift={(9pt,-2pt)}]2-2.south east);
            % % Second 2x2 block
            % \tikz \draw[thick]
            %     ([shift={(-2pt,2pt)}]3-3.north west)
            %     rectangle
            %     ([shift={(2pt,-2pt)}]4-4.south east);
            \end{pNiceMatrix}
            _\mathcal{B},
    \end{equation}
    \begin{equation}
        \hat{L}_1 = 
            \begin{pNiceMatrix}
                1 & \mathbbm{i} & 0 & 0 \\
                \mathbbm{i} & 3 & 0 & 0 \\
                0 & 0 & 1 & \mathbbm{i} \\
                0 & 0 & \mathbbm{i} & 3 \\
                \CodeAfter
                % % First 2x2 block
                % \tikz \draw[thick]
                %     ([shift={(-2pt,2pt)}]1-1.north west)
                %     rectangle
                %     ([shift={(2pt,-2pt)}]2-2.south east);
                % % Second 2x2 block
                % \tikz \draw[thick]
                %     ([shift={(-2pt,2pt)}]3-3.north west)
                %     rectangle
                %     ([shift={(2pt,-2pt)}]4-4.south east);
            \end{pNiceMatrix}
            _\mathcal{B}, \quad
        \hat{L}_2 = 
            \begin{pNiceMatrix}
                1 & 0 & 0 & 0 \\
                \mathbbm{i} & 2 & 0 & 0 \\
                0 & 0 & 1 & 0 \\
                0 & 0 & \mathbbm{i} & 2 \\
                \CodeAfter
                % % First 2x2 block
                % \tikz \draw[thick]
                %     ([shift={(-2pt,2pt)}]1-1.north west)
                %     rectangle
                %     ([shift={(2pt,-2pt)}]2-2.south east);
                % % Second 2x2 block
                % \tikz \draw[thick]
                %     ([shift={(-2pt,2pt)}]3-3.north west)
                %     rectangle
                %     ([shift={(2pt,-2pt)}]4-4.south east);
            \end{pNiceMatrix}
            _\mathcal{B}.
    \end{equation}
    These operators exhibit the block-diagonal structure described in Eq.~\eqref{eq: fragmentation_general}. Using Hamiltonian $\hat{H}$, one finds that 
    \begin{equation}
    \begin{split}
        \bf \Delta_H = & E\ 
            \begin{pmatrix}
                3 & 6 - 2\mathbbm{i} \\
                6 + 2\mathbbm{i} & 3
            \end{pmatrix} 
            - E \ 
            \begin{pmatrix}
                -3 & 6 \\
                6 & -3
            \end{pmatrix} \\
        = & E \ 
            \begin{pmatrix}
                6 & - 2\mathbbm{i} \\
                2\mathbbm{i} & 6
            \end{pmatrix} ,
    \end{split}
    \end{equation}
    which is not proportional to the identity operator. This is in contrast to the previous examples, where the respective $\bf \Delta_H$ is proportional to the identity operator.

    Let us now consider a specific choice of parameters, $E = 1$, $\gamma_1 = 1$, and $\gamma_2 = 8$. The steady mode of $\mathscr{L}^\sharp$ is $ {\bf R_* } = \frac{1}{2} \begin{psmallmatrix} 1 & \mathbbm{i} \\ -\mathbbm{i} & 1 \end{psmallmatrix} $, which satisfies $ {\bf \Delta_H R_* } = 4 {\bf R_*}$. Consequently, this configuration supports a pair of persistent oscillatory modes with an oscillation frequency $\omega = \pm 4$. This behavior is confirmed by the Liouvillian eigenspectrum shown in 
    % Fig.~\ref{fig: example_3_spectrum_osc}, 
    Fig.~\ref{fig: plot_3}(a), 
    where two purely imaginary eigenvalues, $\Lambda = \pm \mathbbm{i} 4$ are shown. It can also be verified that the dissipators corresponding to these two persistent oscillatory modes are not zero.

    Unlike the previous example (Sec. \ref{sec: Example 2}) the persistent oscillatory modes exist independently of the damping rate $\gamma$. The proposed model has fine-tuning sensitivity. Specifically, the presence of oscillatory modes depends critically on the choice of $\gamma_1$ and $\gamma_2$. For instance, if we choose $E = 1$, $\gamma_1 = 1$, and $\gamma_2 = 1$, this only gives a steady mode $\mathscr{L}^\sharp$ where $ {\bf R_* } = \frac{1}{251} \begin{psmallmatrix} 73 & -70 + \mathbbm{i} 52 \\ -70 - \mathbbm{i} 52 & 178 \end{psmallmatrix} $. As the $\bf \Delta_H$ relation is invalidated, the system does not support persistent oscillatory modes. This can be seen in the corresponding Liouvillian eigenspectrum in 
    % Fig.~\ref{fig: example_3_spectrum_damp}, 
    Fig.~\ref{fig: plot_3}(b), 
    where no purely imaginary eigenvalues are found. 
    
% \newpage
\section{\label{sec: Summary} Summary }

    We have proposed a theoretical framework for constructing persistent oscillatory modes in Markovian open quantum systems governed by the time-independent Lindblad master equation. Our framework extends beyond the decoherence-free subspace approach. The primary condition for the existence of such modes is that the Hamiltonian  $\hat{H}$ and the jump operators $\hat{L}_i$ have a block structure given by Eq.~\eqref{eq: fragmentation_general}. Combined with the weak $\bf \Delta_H$ condition in Eq.~\eqref{eq: weak Delta H condition} facilitates the existence of persistent oscillatory mode. However, these modes are fragile and depend sensitively on the parameters. To overcome this, the strong $\bf \Delta_H$ condition in Eq.\eqref{eq: strong Delta H condition} can be used, which eliminates dependence on $\bf R_*$, thereby avoiding fine-tuning. Under these conditions, the system exhibits oscillatory dynamics even in the presence of dissipation, thus generalizing the concept of decoherence-free subspaces to encompass persistent oscillatory modes. Through the illustrative models, we have demonstrated that appropriately engineered system–environment couplings that satisfy these conditions stabilize long-lived coherent oscillations. This framework provides a systematic route for realizing and controlling long-lived coherent dynamics in open quantum systems. It is interesting to explore how our method can be extended to systems with large number of spins, which may provide insights into the understanding of limit cycle phases in thermodynamic limit~\cite{chanLimitcyclePhaseDrivendissipative2015,PhysRevLett.121.035301,bucaNonstationaryCoherentQuantum2019,seiboldDissipativeTimeCrystal2020,bakkerDrivenDissipativeTimeCrystalline2022,kongkhambutObservationContinuousTime2022,wadenpfuhlEmergenceSynchronizationDrivenDissipative2023,dingErgodicityBreakingRydberg2024,wuDissipativeTimeCrystal2024,xiangSelforganizedTimeCrystal2024a,PhysRevLett.134.050407}. 

\textit{Acknowledgments}--We thank Yuqiang Liu for helpful discussion. We acknowledge the support from the EPSRC through Grant No. EP/W015641/1 and No. EP/W524402/1. The data that support the findings of this study are openly available~\cite{Zenodo}.

% \clearpage

%%\nocite{*}

\bibliography{citation}% Produces the bibliography via BibTeX.

%=============================================================================
\end{document}